\documentclass[%
 reprint,%
 amssymb, amsmath,%
 aip,jcp,%
]{revtex4-1}

\usepackage{graphicx}
\usepackage{docs}%
\usepackage{bm}%
\usepackage[colorlinks=true,linkcolor=blue,breaklinks]{hyperref}%
\expandafter\ifx\csname package@font\endcsname\relax\else
 \expandafter\expandafter
 \expandafter\usepackage
 \expandafter\expandafter
 \expandafter{\csname package@font\endcsname}%
\fi
\newcommand{\imi}{\mathrm{i}}

\newcommand{\kets}[1]{\left|#1\right\rangle}
\newcommand{\bras}[1]{\left\langle#1\right|}
\newcommand{\inprdt}[2]{\left\langle #1 \right.\left| #2 \right\rangle}

\begin{document}

\title{Decoherence and Energy Relaxation in the Quantum-Classical Dynamics for Charge Transport in Organic Semiconducting Crystals: an Instantaneous Decoherence Correction Approach}%

\author{Wei Si}\affiliation{State Key Laboratory of Surface Physics and Department of Physics, Fudan University, Shanghai 200433, China}
\author{Chang-Qin Wu}\email[Email: ] {cqw@fudan.edu.cn} \affiliation{State Key Laboratory of Surface Physics and Department of Physics, Fudan University, Shanghai 200433, China} \affiliation{Collaborative Innovation Center of Advanced Microstructures, Fudan University, Shanghai 200433, China}

\date{\today}%
\revised{}%

\begin{abstract}
We explore an instantaneous decoherence correction (IDC) approach for the decoherence and energy relaxation in the quantum-classical dynamics of charge transport in organic semiconducting crystals. These effects, originating from environmental fluctuations, are essential ingredients of the carrier dynamics. The IDC is carried out by measurement-like operations in the adiabatic representation. While decoherence is inherent in the IDC, energy relaxation is taken into account by considering the detailed balance through the introduction of energy-dependent reweighing factors, which could be either Boltzmann (IDC-BM) or Miller-Abrahams (IDC-MA) type. For a non-diagonal electron-phonon coupling model, it is shown that the IDC tends to enhance diffusion while energy relaxation weakens this enhancement. As expected, both the IDC-BM and IDC-MA achieve a near-equilibrium distribution at finite temperatures in the diffusion process, while the Ehrenfest dynamics renders system tending to infinite temperature limit. The resulting energy relaxation times with the two kinds of factors lie in different regimes and exhibit different dependence on temperature, decoherence time and electron-phonon coupling strength, due to different dominant relaxation process.
\end{abstract}

\maketitle

\section{Introduction}
The advances of organic field-effect devices in recent years have inspired renewed research interest in the charge transport problems in organic semiconducting crystals.\cite{review_ofet, bredas_review} Recent studies focus on the intrinsic regime of charge transport, featured by observation of both band-like and hopping-like properties in the same device under different operating conditions.\cite{hist1, hist2, hall1, hall2, spectro_nm} Besides, the underlying electronic states are proved to be localized both theoretically\cite{mean_free_path} and experimentally. \cite{local_prb, local_prl} It is realized that the coupling between carriers, which is denoted as electrons for simplicity throughout the paper, and lattice vibrations (phonons) plays an essential role in the charge transport process. Further, in many cases, the energy scales of the electron-phonon coupling, temperature and intermolecular transfer integral overlap, which renders the problem in the intermediated parameter regime, where the traditional pictures may not be effective.\cite{wang_sh} For example, Troisi and Orlandi proposed that the carriers in such materials are no longer localized by the self-trapping mechanism in polaron problems.\cite{yu_breather} Rather, it is due to dynamic disorder brought by phonons at finite temperature. Many theoretical works have been contributed to this topic.\cite{polaron_bobbert, polaron_silbey1, polaron_silbey2, green_fratini, green_fratini2, hsr_cao, troisi_dd, tl_fratini, li_symmetry, wang_sh, sse_zhao, zhao_sse2, heom_shi, nt_shuai, coherence_mobility, yao_idc, filippis_ssh} Among them, the series of studies using quantum-classical dynamics, which treat the electronic system quantum-mechanically and the phonons classically, are promising ones. \cite{troisi_dd, wang_sh} They offer a time-dependent view of the transport process, enabling the calculation of quantities such as the transient conductivity. \cite{tl_fratini} Computationally, they are more efficient than the full quantum-mechanical methods and are convenient to be tuned for realistic materials. For example, the detailed crystal and electronic structures extracted by experiments and first-principle calculations can be input as model parameters for multi-scale simulations.\cite{troisi_prl2} Besides, existing studies on chemical dynamics provide valuable resources for the improvement of those methods. For example, Wang et al. successfully generalized the surface hopping approach to such systems,\cite{wang_sh} revealing an interesting crossover between band-like and hopping behavior.

Despite the successes of the quantum-classical dynamics, there are still some inconsistencies in the resulting carrier dynamics. For example, in the pioneering work of Troisi et al. by Ehrenfest dynamics,\cite{troisi_dd} the form of evolution of the electronic system is quantum-mechanically coherent. The electron wave functions keep expanding without a well-defined localization length, which does not coincide with experimental findings.\cite{local_prb,local_prl}. This could be ascribed to the lack of explicit treatment of decoherence.\cite{sh_sb,coherence_mobility, deco_assist} In the charge transport process, the electronic system (electron) can be seen as the central system. In realistic situations, it would interact with many other species (environment), especially the dominant phonon degrees of freedom that are treated classically here. Together, they form a system-environment architecture. One direct consequence for the system from coupling with environment is the destruction of phase coherence within the system, \textit{i.e.}, decoherence,\cite{sergei_sh} which is essential for the consistency of quantum-classical dynamics.\cite{prezhdo_dish} The incorporation of decoherence is indicated or proposed in several further studies. Fratini et al. introduced a relaxation time approximation, in which the information of initial dynamics are used and the long-time dynamics are corrected.\cite{tl_fratini} Yao et al. introduced the instantaneous decoherence correction (IDC) by measurement-like operations in the lattice site representation, through which band-like temperature dependence of mobility and localized electronic states coexist.\cite{yao_idc} Also, in the surface hopping studies of Wang et al., decoherence corrections are included by collapsing the electronic states to the active states at the surface hopping events. \cite{wang_sh}

Besides decoherence, another prominent process due to the system-environment coupling is the energy relaxation (dissipation),\cite{trugman_relaxation} by which the system absorbs/dissipates energy from/to the environment and relaxes to the thermal equilibrium with the environmental temperature.\cite{zhao_sse2} This process is essential for the interpretation of transient experiments, such as the pump-probe ones.\cite{swanson_pumpprobe,zhao_relaxation} However, there are signs that it is not treated properly in some forms of quantum-classical dynamics. For example, if the effect of phonon motion is taken to be a stochastic potential for the electronic system, all the adiabatic states tend to be populated with equal probability in the long time limit.\cite{hsr_cao, sse_zhao, zhao_sse2} This means the effective temperature of the electronic system is infinitely large. Thus these methods are rigourous only when the thermal energy is much larger than the bandwidth. In the Ehrenfest dynamics the electronic system could influence the phonons by the mean-field Hellmann-Feynman force. However, it does not prevent the electronic system to deviate from the near-equilibrium distribution in the diffusion dynamics,\cite{tl_fratini, tully_ehrenfest} which is also shown in the following. Besides, in the study using Kubo formalism with adiabatic approximation, the dc conductivity turns out to be zero.\cite{filippis_adiabatic} The surface hopping approach is promising in treating the energy relaxation through the frustrated hops in the switching procedure.\cite{tully_balance,tully_balance2,landry_balance} However, these hops also lead to inconsistency problems, the solution of which is under active studies.\cite{prezhdo_dish} Thus the problem of properly accounting for both decoherence and energy relaxation in the quantum-classical dynamics of the dynamic disordered systems remains an open question.

In this paper, we extend the IDC approach \cite{yao_idc} to the adiabatic representation to investigate the decoherence and energy relaxation processes. Similar approaches have been employed to the study of spin dynamics of excited states\cite{measurement2} and carrier dynamics of the Anderson model.\cite{measure_anderson} The evolution of the system in short time regime (comparable to the phonon frequency) are governed by the quantum-classical dynamics. The long time dynamics, which is recognized to be problematic,\cite{tl_fratini} are modified by decoherence corrections. The IDC are carried out by measurement-like operations with different schemes. For energy relaxation, the detailed balance is considered by introducing the IDC with energy-dependent reweighing factors, such as the Boltzmann (IDC-BM) and the Miller-Abrahams (IDC-MA) factors. The IDC with only destruction of phase coherence (IDC-DP) is also studied for comparison. Based on the off-diagonal electron-phonon coupling model, the physical consequences of the IDC approach are explored. This paper is organized as follows. In Section II, the model and the IDC approach are presented in detail. In Section III, the results of the IDC are shown, including both diffusion and electronic energy. The paper is summarized briefly in Section IV.

\section{Model and Method}
In this section, we first present the off-diagonal electron-phonon coupling model and the equations of motion. We then move on to introduce the IDC approach with a couple of schemes for energy relaxation, including that with Boltzmann and Miller-Abrahams factors.
\subsection{Model}
The Hamiltonian we consider in this work is composed of the electronic and the phonon part, which is $H=H_{\mathrm{el}}+H_{\mathrm{ph}}$.\cite{troisi_dd,ssh_review} The electronic part is
\begin{equation}
H_{\mathrm{el}} = \sum_j J[-1+\alpha(u_{j+1}-u_j)](c^{\dagger}_jc_{j+1}+c^{\dagger}_{j+1}c_j),
\end{equation}
where $J$ is the transfer integral, $\alpha$ the electron-phonon coupling constant, $u_j$ the displacement of phonon on the $j$-th site, $c^{\dagger}_j$ ($c_j$) the creation (annihilation) operators of electron. The Hamiltonian for the phonons is
\begin{equation}
H_{\mathrm{ph}} = \sum_j\left[\frac{1}{2}m\dot{u}_j^2+\frac{1}{2}m\omega_0^2u_j^2\right],
\end{equation}
where $m$ is the effective mass of the phonon and $\omega_0$ is the frequency. The electron state is described by the wave function $\kets{\psi(t)}$ that is governed by the Schr\"{o}dinger equation, which is
\begin{equation}
\imi\hbar\frac{\partial \kets{\psi(t)}}{\partial t}=H_{\mathrm{el}}\kets{\psi(t)}.
\end{equation}
The electronic system back-reacts on the phonons through the mean-field Hellmann-Feynman force. The Langevin heat bath is included to account for the fluctuations of the phonons due to the thermal environment. The equations of motion for the phonons is
\begin{equation}
m\ddot{u}_j=-m\omega_0^2u_j-\frac{\partial E_{\mathrm{el}}}{\partial u_j}-\gamma m\dot{u}_j+\xi_j,
\end{equation}
where $E_{\mathrm{el}}=\bras{\psi(t)}H_{\mathrm{el}}\kets{\psi(t)}$ is the expectation value of energy at time $t$, $\gamma$ the friction constant of the Langevin heat bath and $\xi_j$ the stochastic forces. $\xi_j$ satisfy the correlation function $\langle\xi_i(t)\xi_j(0)\rangle=\sqrt{2\gamma m k_B T\delta(t)}\delta_{ij}$, where $k_B$ is the Boltzmann constant and $T$ is the temperature. The initial displacements $u_j(0)$ and velocities $\dot{u}_j(0)$ are drawn from the Maxwell distributions with variance $k_B T/m\omega_0^2$ and $k_B T/m$ respectively. The initial state of the electron is chosen among the adiabatic states (energy eigenstates) at $t=0$ according to the Boltzmann distribution $P_{\nu}=\exp(-E_{\nu}/k_B T)/\sum_{\mu}\exp(-E_{\mu}/k_B T)$. The evolution is carried out by the $4$th order Runge-Kutta method and the stochastic forces are incorporated by the method of Wang et al.\cite{wang_sh,langevin} The final result is averaged over enough realizations for the convergence of relevant physical quantities.

\subsection{Instantaneous Decoherence Correction}
We now move on to describe the instantaneous decoherence corrections (IDC), which is carried out by repeated measurement-like operations on the electronic wave function in the adiabatic representation.\cite{yao_idc} A decoherence time $t_d$ is introduced as the time interval between successive IDC. It reflects the decoherence rate due to the system-environment interaction. This approach is shown to be effective to incorporate decoherence in the dynamics.\cite{wang_sh, sergei_sh} Although discrete, it is shown in the site representation that the results resemble those by actually adding an attenuation term in the equation of motion of the non-diagonal elements of the density matrix.\cite{yao_idc} For the following results, a fixed $t_d$ is used for convenience. Actually, more elaborate forms for $t_d$ can be taken to reflect the situation in realistic materials. For example, proper distributions, such as the Poisson distribution,\cite{prezhdo_dish} can be used, together with specific models to determine $t_d$ from other parameters.\cite{bittner_decoheretime, truhlar_csdm,subotnik_decoheretime}

In detail, the dynamics are sliced into segments of time period $t_d$. Within each $t_d$, the carrier and phonons are evolved by the quantum-classical dynamics. After each $t_d$, a measurement-like operation is carried out for the electronic system by collapsing the electron state to a certain adiabatic state according to a chosen distribution, which can be set by different schemes. Suppose the adiabatic states are denoted as $\kets{E_{\mu}}$ with energy $E_{\mu}$ and the distribution is $P_{\mu}$, where $\mu$ is the index. A random number $\chi$ is chosen uniformly in $[0,1)$. The wave function is to be reset to that of $\kets{E_{\nu}}$ if
\begin{equation}
\sum_{\mu=0}^{\nu-1} P_{\mu} \leq \chi < \sum_{\mu=0}^{\nu} P_{\mu}.
\end{equation}

For the distribution, the direct choice is that with the wave function in the adiabatic representation, which is
\begin{equation}
P_{\mu}=|\inprdt{E_{\mu}}{\psi(t)}|^2,
\end{equation}
where $\kets{\psi(t)}$ is the wave function prior to the decoherence correction at time $t$. This scheme removes the phase relation among adiabatic states and is termed as the IDC with destruction of phase coherence (IDC-DP). However, as is shown below, this scheme only includes decoherence and energy relaxation is still not properly treated.

\subsection{Energy Relaxation}
For the purpose of energy relaxation in the IDC, we consider the detailed balance through the introduction of energy-dependent reweighing factors. We propose a couple of schemes in this work. The first one is to implement a Boltzmann factor, which is
\begin{equation}
P_{\mu}^{\text{BM}}=P_{\mu}\cdot\exp(-E_{\mu}/k_BT)/C_{\text{BM}},
\end{equation}
where $C_{\text{BM}}$ is the normalization factor. This scheme could reflect the ability of the environment to recover the thermal distribution for the electronic system and is denoted as the IDC-BM in the following. It is motivated by Einstein's theory of spontaneous emission of excited states,\cite{einstein} which is the result of quantum fluctuations of radiation fields. Further, motivated by the Miller-Abrahams formula,\cite{ma_formula} an alternative scheme is implemented. If the measurement is to collapse the state to an adiabatic state with higher energy, the probability is reduced by a factor from actually absorbing the energy from the environment. In this sense, the distribution is
\begin{equation}
P_{\mu}^{\text{MA}}=\left\{
\begin{array}{ll}
P_{\mu}\exp[-(E_{\mu}-\bar{E})/k_BT]/C_{\text{MA}}, & E_{\mu}>\bar{E}, \\
P_{\mu}/C_{\text{MA}}, & E_{\mu}\leq\bar{E},
\end{array}
\right.
\end{equation}
where $C_{\text{MA}}$ is a normalization constant. This scheme is termed as the IDC-MA in the following. Similar expressions are widely used for the hopping rates in the simulation of charge transport in amorphous organic semiconductors.\cite{bredas_review} They are shown to capture the transient mobility behavior\cite{bobbert_relaxation} and charge extraction transients in organic solar cells,\cite{kemerink_relaxation} which depend on the correct treatment of energy relaxation.

It is noted that with the IDC, the electronic states remain fairly localized. It is not necessary to solve the Hamiltonian of the whole lattice and a specific region is chosen instead, within which the electronic state is populated, which is more efficient. The region is determined as follows. The index of the left/right boundary of the region is denoted as $j_{l/r}$. The quantity $p_{l/r}=\sum_{\mu} P_{\mu}|\inprdt{j_{l/r}}{\mu}|^2$ is calculated and is ensured to be smaller than a critical value. Otherwise the region should be expanded. Actually, $p_{l/r}$ reflect the expectation value of populations on the boundary sites after the decoherence correction. In the following calculations, the critical value is chosen to be $10^{-6}$, which is small enough to not influence the final result.

\section{Results}
For the following results, the parameters are taken to be typical for pentacene.\cite{troisi_dd} The transfer integral $J$ is $0.037\text{ }e$V, the electron-phonon coupling constant $\alpha=3.3$ {\AA}$^{-1}$, the phonon frequency $\omega_0=7.6$ ps$^{-1}$, the phonon effective mass $m=250$ amu; the lattice constant $a=4$ {\AA}; the temperature $T=150$ K. The friction constant $\gamma$ is taken to be $1$ ps$^{-1}$. \cite{wang_sh} A lattice with $600$ sites is taken to avoid the boundary effects. The decoherence time is taken to be $t_d=10\hbar/J\approx 180$ fs. This set of parameters is taken in the following unless stated otherwise.

\subsection{Diffusion}
We first consider the diffusion with the IDC, which is reflected by the time-dependent averaged population of electron among lattice sites
\begin{equation}
P_j(t)=\frac{1}{N_s}\sum_{s}\left|\psi_j^s(t)\right|^2,
\end{equation}
where $\psi_j^s(t)$ the coefficient of the electron wave function for the $j$-th site in the $s$-th realization. $N_s$ is the total number of realizations, which is taken to be at least $10000$. The mean squared displacement (MSD) is calculated from the distribution $P_j(t)$, which is
\begin{equation}
\mathrm{MSD}(t)=a^2\left\{\sum_j j^2P_j(t)-\left[\sum_jjP_j(t)\right]^2\right\}.
\end{equation}
The diffusion constant is the time derivative of $\mathrm{MSD}$ in the long time limit, which is
\begin{equation}
D=\frac{1}{2}\lim_{t\rightarrow\infty}\frac{\mathrm{d}\mathrm{MSD}(t)}{\mathrm{d}t}.
\end{equation}
The evolution time is taken to be at least $10$ ps for the convergence of $D$.

\begin{figure}
\includegraphics{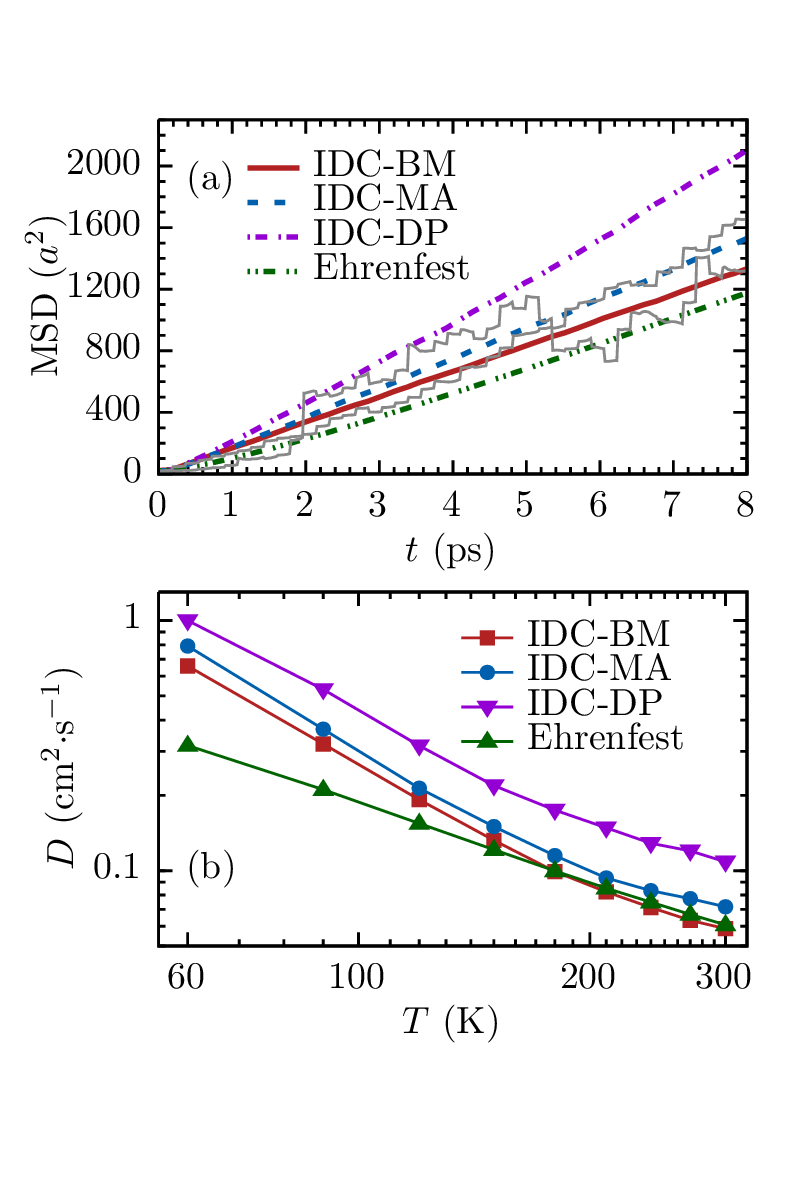}
\caption{\label{fig01} (a) Evolution of the mean squared displacement (MSD) with different instantaneous decoherence correction (IDC) schemes applied to an off-diagonal electron-phonon coupling model, in comparison with the result of Troisi's method with Ehrenfest dynamics (green dot-dot-dash): IDC with Boltzmann factor (IDC-BM, red solid), IDC with Miller-Abrahams factor (IDC-MA, blue dash) and IDC with destruction of phase coherence (IDC-DP, purple dot-dash). All results are got by averaging over more than 10000 realizations. The results with the IDC-BM with 10 and 100 realizations are also shown (grey solid). (b) Temperature dependence of diffusion constant with the above IDC schemes: IDC-BM (red square), IDC-MA (blue circle), IDC-DP (purple lower-triangle) and Ehrenfest dynamics (green upper triangle).}
\end{figure}

The typical MSD with the above IDC schemes are shown in Fig. \ref{fig01}(a), together with that from the Ehrenfest dynamics of Troisi et al.\cite{troisi_dd}
The prominent effect of the IDC in adiabatic representation is that the diffusion is enhanced, in contrast to the suppression of diffusion with the IDC in site representation,\cite{yao_idc} which is a result of the quantum Zeno effect.\cite{zeno} The enhancement is reduced by inclusion of energy relaxation with the IDC-BM and IDC-MA schemes. For comparison, the results after averaging over 10 and 100 realizations with the IDC-BM are also shown. It can be seen that with 100 realizations, the result is already close to the final one.

The enhancement of diffusion originates from the way the electron diffuses in the quantum-classical dynamics. Take the IDC-DP as an example. The average population on the $j$-th site before the decoherence correction is
\begin{equation}
P_j^b=|\sum_{\mu}\inprdt{j}{E_{\mu}}\inprdt{E_{\mu}}{\psi}|^2.
\end{equation}
The decoherence correction destroys the phase coherence between different adiabatic states and the corresponding population after the decoherence correction is
\begin{equation}
P_j^a=\sum_{\mu}|\inprdt{j}{E_{\mu}}|^2|\inprdt{E_{\mu}}{\psi}|^2.
\end{equation}
Comparing with $P_j^b$, the missing terms in $P_j^a$ are the interference ones between adiabatic states. In the Ehrenfest dynamics, the wave function starts to change by the mixing of adiabatic states due to phonon motion. If the wave function is expanded in the evolved adiabatic states, the interference terms tend to be constructive for those sites that are initially populated and destructive for those sites that are initially unpopulated. With further evolution, this kind of phase relation is disrupted by the stochastic phonon motion and population emerges on those initially unpopulated sites. The IDC helps this process and thus enhances diffusion. A specific example of this enhancement is shown in the supplementary material. \cite{supple}

We further calculate the temperature dependence of diffusion constants. The results are shown in Fig. \ref{fig01}(b). It can be seen that all the IDC schemes, like the Ehrenfest dynamics, give a band-like dependence of diffusion constant. In the lower temperature regime around $100$ K, the diffusion constant with the IDC are substantially larger than the ones given by the Ehrenfest dynamics. With increasing temperature, the diffusion constants from the IDC-DP scheme remain higher than the Ehrenfest ones, while those from the IDC-BM and IDC-MA schemes gradually decrease and nearly coincide with the Ehrenfest ones. Besides, the diffusion constants with the IDC show sharper dependence on temperature. Beyond $150$ K, they begin to deviate from a power-law behavior. This can be related to the corresponding deviation of the localization length of adiabatic states. The temperature dependence of the averaged localization length is provided in the supplementary material.\cite{supple}

\subsection{Energy Distribution of Electronic States}
\begin{figure}
\includegraphics{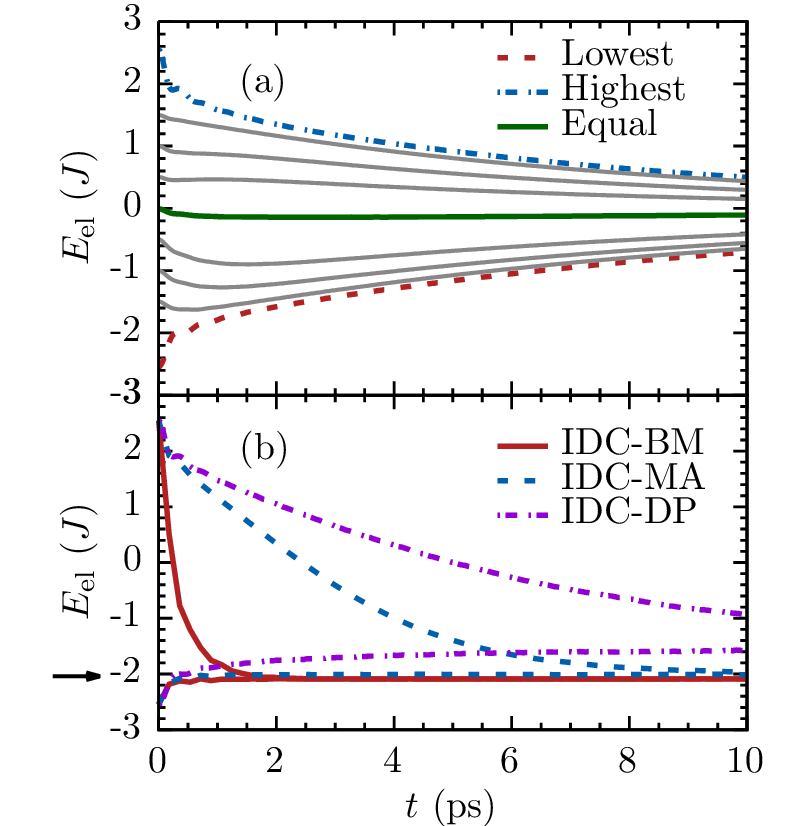}
\caption{\label{fig02} (a) Evolution of the electronic energy $E_{\mathrm{el}}$ with the Ehrenfest dynamics starting from the adiabatic states with lowest (red dash) and highest (blue dot-dash) energies, and energies that are closest to $\pm 1.5J$, $\pm J$, $\pm 0.5J$ (grey solid). The result starting from random adiabatic states with equal probability (green solid) is also shown. (b) Evolution of $E_{\mathrm{el}}$ for the IDC-BM (red solid), IDC-MA (blue dash) and IDC-DP (purple dot-dash) schemes starting from adiabatic states with the lowest and the highest energies. The average value of the Boltzmann distribution is shown by the arrow on the left.}
\end{figure}

We now turn our attention to the electronic energy $E_{\mathrm{el}}=\bras{\psi(t)}H_{\text{el}}\kets{\psi(t)}$ for the energy relaxation process. Firstly, the evolution of $E_{\mathrm{el}}$ in the Ehrenfest dynamics are shown in Fig. \ref{fig02}(a).
The initial condition of the electronic system is chosen to be adiabatic states with both the highest and the lowest energies, and those closest to $\pm 1.5J$, $\pm J$, $\pm 0.5J$. They all tend to relax to energy values near the center of the density of states ($E=0$). To check the observation, the evolution of $E_{\mathrm{el}}$ starting from random adiabatic states with equal probability is also shown. An initial small decrease away from zero is observed, which persists beyond $10$ ps. The small deviation is caused by the polaron effect due to the Hellmann-Feynman force, which is a feedback from the electronic system to the phonons. It is confirmed by the evolution of averaged displacements and phonon potential energies that are shown in the supplementary material.\cite{supple} These results show that the near-equilibrium Boltzmann distribution is violated in the Ehrenfest dynamics. All the adiabatic states tend to be populated with equal probability, which corresponds to a large effective temperature. This violation is from the inability of the coherent quantum dynamics to distinguish between the processes that go upward and downward in energy. Similar problems are also pointed out in the studies of the spin-boson dynamics by stochastic Sch\"{o}dinger equation with real-valued stochastic terms \cite{sse_zhao} and the Anderson model with Haken-Strobl-Reineker method.\cite{hsr_cao}

We then study the evolution of the electronic energy with the IDC schemes starting from adiabatic states with both the highest and the lowest energies, which are shown in Fig. \ref{fig02}(b). For the IDC-DP scheme, the final energy tends to a finite negative value different from the value of the Boltzmann distribution, which is indicated by the arrow on the left. This negative value has the same origin with the negative dip in the Ehrenfest dynamics as discussed above. Here, the effect is made more pronounced by the decoherence corrections to adiabatic states. Further, it is clear that for the IDC-BM scheme, the electronic energies starting from both the lowest and highest adiabatic states tend to the same limit, which is only slightly lower ($\sim 0.06J$) than the Boltzmann value. For the IDC-MA scheme, the energies also relax to a final value, which is slightly higher than the Boltzmann one. Besides, the IDC-BM scheme gives a much faster relaxation process than the IDC-MA. In all, both schemes help recover the near-equilibrium Boltzmann distribution in the dynamics, which is not observed in the Ehrenfest dynamics and the IDC-DP scheme. This property is present in the temperature regime considered in this paper, which is shown in the supplementary material.\cite{supple}

\begin{figure}
\includegraphics{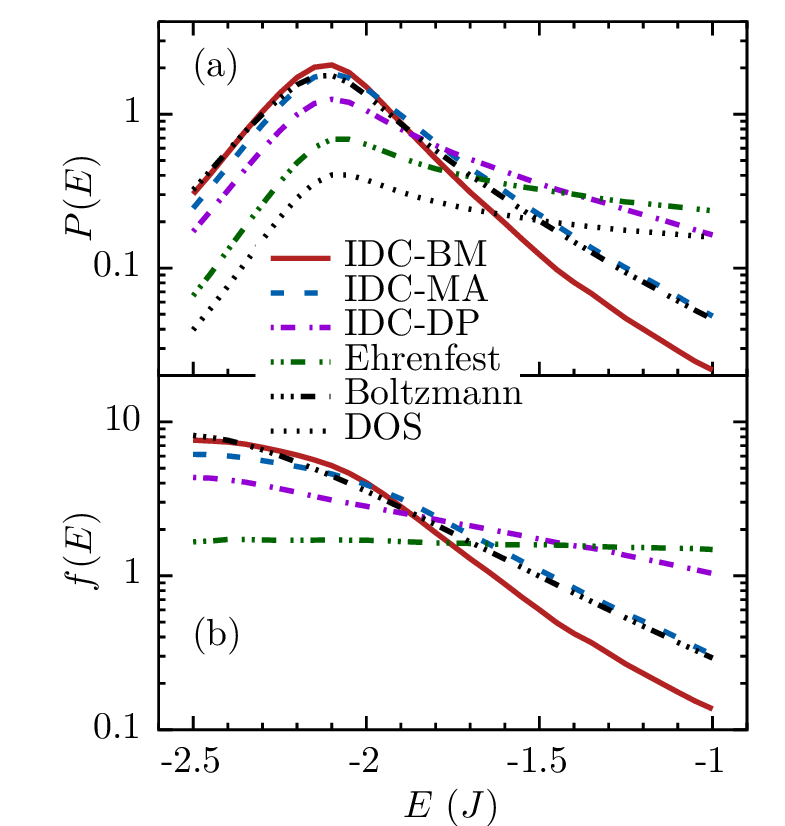}
\caption{\label{fig03}  (a) Population distribution $P(E)$ of the electronic states with respect to energy $E$ after evolution of $10$ ps starting from adiabatic states with the Boltzmann distribution: IDC-BM (red solid), IDC-MA (blue dash), IDC-DP (purple dot-dash) and Ehrenfest dynamics (green dot-dot-dash). The result of the Boltzmann distribution (black triple-dot-dash) and the density of states $D(E)$ (black dot) are also shown for comparison. (b) Corresponding probability distribution defined as $f(E)=P(E)/D(E)$.}
\end{figure}

We further calculated the population distributions $P(E)$ of the the electronic states with respect to energy after evolving the system for $10$ ps, which are shown in Fig. \ref{fig03}(a).
In the calculation of $P(E)$, the contribution from each realization is broadened by a Lorentzian lineshape with width $\Gamma$, which is
\begin{equation}
P(E)=C_{P}\sum_{s,\mu} |\langle E_{\mu}^s|\psi^s\rangle|^2\frac{\Gamma^2}{(E-E_{\mu}^s)^2+\Gamma^2},
\end{equation}
where $C_P$ is the normalization constant to ensure $\int P(E)\mathrm{d}E=1$. Here $\Gamma$ is chosen to be $0.05J$. The density of states $D(E)$ can be defined similarly. A quantity $f(E)$ is further defined to depict the probability distribution of the electronic system as $f(E)=P(E)/D(E)$, which is shown in Fig. \ref{fig03}(b). In both figures, the energy is from $-2.5J$ to $-J$, in which the population is pronounced. A plot over the full energy regime is provided in the supplementary material for clarity.\cite{supple} For comparison, the results of the Boltzmann distribution are also shown. The deviation of the Boltzmann result from linear behavior around $E=-2.2J$ in Fig. \ref{fig03}(b) is an artifact from the fast variation of $D(E)$. In Fig. \ref{fig03}(a), $D(E)$ is also shown as the case when all the adiabatic states are populated with equal probability. It can be seen that the results of the Ehrenfest dynamics tend to $D(E)$ and the results of the IDC-DP scheme also deviate from the near-equilibrium Boltzmann distribution. In contrast, the results of both the IDC-BM and IDC-MA are not far away from equilibrium ones. Besides, in the lower energy regime below the peak of the population, the IDC-BM gives a distribution that is closer to the Boltzmann one; in the higher energy regime, the correspondence is better with the IDC-MA scheme.

\subsection{Energy Relaxation Time}
\begin{figure}
\includegraphics{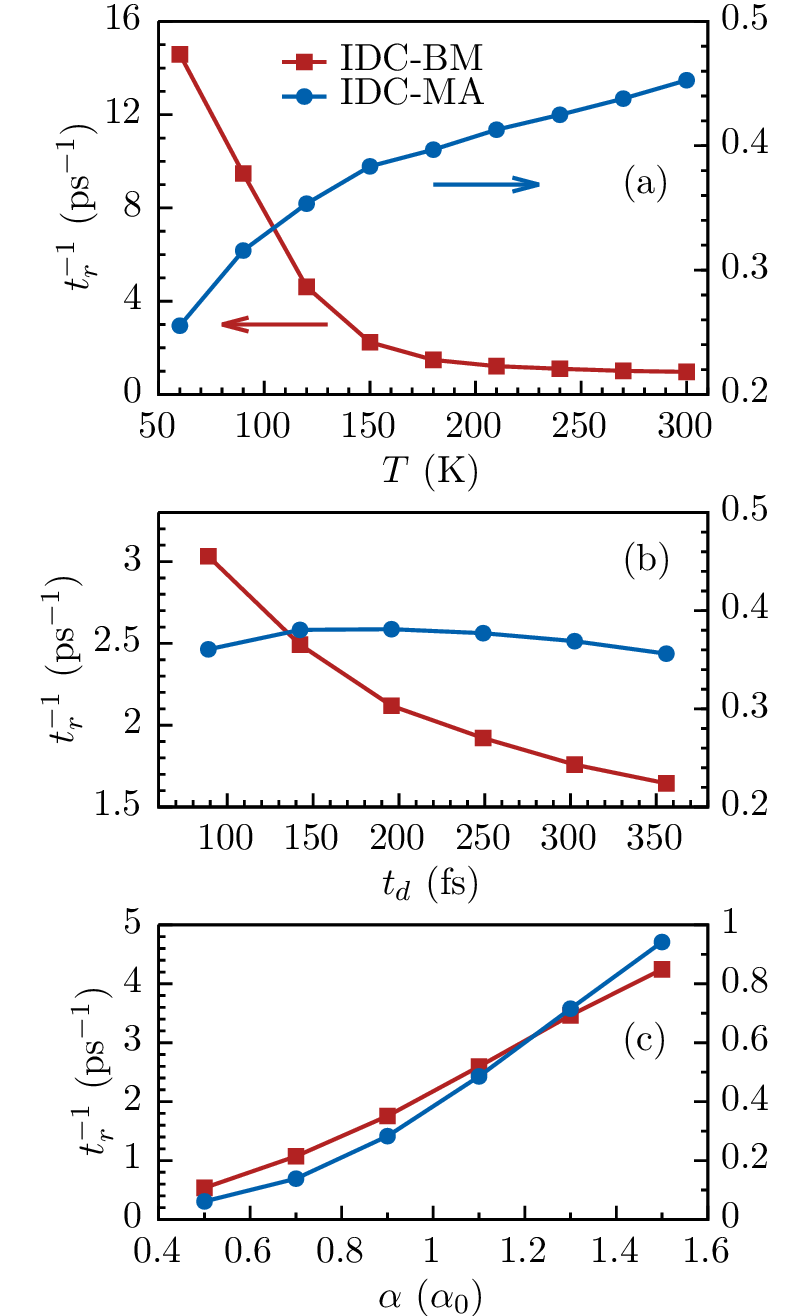}
\caption{\label{fig04} (a) Dependence of the inverse energy relaxation time $t_r^{-1}$ on temperature $T$ for the IDC-BM (red square) and the IDC-MA (blue circle) schemes. The decoherence time is chosen to be $t_d=180$ fs. (b) Dependence of $t_r^{-1}$ with decoherence time $t_d$. The temperature is fixed at $T=150$ K. (c) Dependence of $t_r^{-1}$ with electron-phonon couplings $\alpha$, with $\alpha_0=3.3$ \AA$^{-1}$.}
\end{figure}

Furthermore, we note that an energy relaxation time $t_r$ can be extracted from the the evolution of electronic energy, which is a quantity commonly used for describing the associated transient properties.\cite{trugman_relaxation} For this purpose, we calculate the evolution of $E_{\mathrm{el}}$ from randomly chosen adiabatic states with equal probability. The evolution can be well fitted by an exponential damping function $\sim \exp(-t/t_r)$ in the temperature regions considered. The dependence of the inverse energy relaxation time $t_r^{-1}$ on temperature, decoherence time and electron-phonon couplings are shown in Fig. \ref{fig04}, where large $t_r^{-1}$ means faster energy relaxation.
It can be seen that the behaviors of $t_r^{-1}$ with the IDC-BM and IDC-BM schemes are very different. For the temperature dependence, $t_r^{-1}$ from the IDC-BM scheme decreases with temperature while that from the IDC-MA scheme increases. For the decoherence-time dependence, $t_r^{-1}$ from the IDC-BM scheme decreases with $t_d$ while that from the IDC-MA scheme is insensitive. Both schemes give inverse $t_r^{-1}$ that increase with increasing electron-phonon coupling $\alpha$. These differences come from the different dominant energy relaxation processes. In the IDC-BM, energy relaxation is dominated by the decoherence correction. With lower temperature, the Boltzmann factor $\sim\exp(-E/k_BT)$ has greater ability to redistribute the population toward lower adiabatic states. For smaller $t_d$, the redistribution is more frequent. In both cases, the relaxation is faster. In contrast, for the IDC-MA scheme, energy relaxation is through the energy fluctuation of adiabatic states. The decoherence corrections bring about the imbalance between the processes that go upward and downward in energy. With lower temperature, the fluctuation is slower, leading to slower energy relaxation. Besides, different $t_d$ do not influence the fluctuation substantially, thus leading to the insensitivity. For both schemes, increasing electron-phonon coupling leads to faster mixing of adiabatic states and faster energy relaxation. Besides, it is noted that $t_r$ with the IDC-MA ($1$ ps) are generally ne order of magnitude larger than the implemented decoherence time $t_d$ ($100$ fs), which follow the same relationship revealed in the studies of some elementary models, such as the spin-boson model. \cite{sb_review} In contrast, $t_r$ with the IDC-BM are comparable to $t_d$, which may correspond to a faster relaxation mechanism.\cite{zhao_relaxation}

\section{Summary}
In summary, we have studied the decoherence and energy relaxation, which are essential processes induced by the (quantum) system-environment interaction in the quantum-classical dynamics of charge transport in organic semiconducting crystals. They can be incorporated by the IDC with either the Boltzmann or the Miller-Abrahams factors. We find that, while decoherence enhances diffusion, the energy relaxation weakens this enhancement. With both the IDC-BM and IDC-MA, the distributions of electronic states tend to the near-equilibrium one, which is a sign of proper treatment of energy relaxation. The energy relaxation time from the two schemes show different behaviors in its dependence on temperature, decoherence time and electron-phonon coupling strength, which result from different relaxation processes. Furthermore, it can be shown that the decoherence and energy relaxation are crucial for observing a direct current response with quantum-classical dynamics, which make a direct calculation of mobility under a finite external electric field possible.\cite{dong_field}

\begin{acknowledgments}
W. Si would like to thank Jing-Juan Dong for her helpful discussions. We acknowledge the financial supports from the National Natural Science Foundation of China and the National Basic Research Program of China (2012CB921401).
\end{acknowledgments}

\end{document}